# Valley-dependent Exciton Fine Structure and Autler-Townes Doublets from Berry Phases in Monolayer MoSe$_2$

Chaw-Keong Yong[†*,1], M. Iqbal Bakti Utama[†,1,2,8], Chin Shen Ong[1,8], Ting Cao[1,8], Emma C. Regan[1,3,8], Jason Horng[4], Yuxia Shen[5], Hui Cai[6], Kenji Watanabe[7], Takashi Taniguchi[7], Sefaattin Tongay[5], Hui Deng[4], Alex Zettl[1,8,9], Steven G. Louie[1,8], Feng Wang[*,1,8,9]

[1] Department of Physics, University of California at Berkeley, Berkeley, California 94720, United States.

[2] Department of Materials Science and Engineering, University of California at Berkeley, Berkeley, California 94720, United States.

[3] Graduate Group in Applied Science and Technology, University of California at Berkeley, Berkeley, California 94720, United States.

[4] Department of Physics, University of Michigan, 450 Church Street, Ann Arbor Michigan 48109-2122, United States.

[5] School for Engineering of Matter, Transport and Energy, Arizona State University, Tempe, Arizona 85287, United States.

[6] Center for Nanophase Materials Sciences, Oak Ridge National Laboratory, Oak Ridge, Tennessee 37831, United States.

[7] National Institute for Materials Science, 1-1 Namiki, Tsukuba, 305-0044, Japan.

[8] Materials Sciences Division, Lawrence Berkeley National Laboratory, Berkeley, California 94720, United States.

[9] Kavli Energy NanoScience Institute at University of California Berkeley, Berkeley, California 94720, United States.

† These authors contributed equally to this work.

* Correspondence to: fengwang76@berkeley.edu, chawkeong@berkeley.edu.

## Abstract

The Berry phase of Bloch states can have profound effects on electron dynamics[1-3] and lead to novel transport phenomena, such as the anomalous Hall effect and the valley Hall effect[4-6]. Recently, it was predicted that the Berry phase effect can also modify the exciton states in transition metal dichalcogenide monolayers, and lift the energy degeneracy of exciton states with opposite angular momentum through an effective valley-orbital coupling[1,7-11]. Here, we report the first observation and control of the Berry-phase induced splitting of the 2p-exciton states in monolayer molybdenum diselenide (MoSe$_2$) using the intraexciton optical Stark spectroscopy. We observe the time-reversal-symmetric analog of the orbital Zeeman effect resulting from the valley-dependent Berry phase, which leads to energy difference of +14 (-14) meV between the 2p$_+$ and 2p$_-$ exciton states in K (K') valley, consistent with the ordering from our *ab initio* GW-BSE results. In addition, we show that the light-matter coupling between intraexciton states are remarkably strong, leading to prominent valley-dependent Autler-Townes doublet under resonant driving. Our study opens up new pathways to coherently manipulate the quantum states and excitonic excitation with infrared radiation in two-dimensional semiconductors.

In the momentum space of atomically thin transition metal dichalcogenides (TMDs), a pair of degenerate exciton states are present at the K and K'-valleys, producing a valley degree of freedom that is analogous to the electron spin[12-14]. The electrons in the K and K'-valleys acquire a finite Berry phase when they traverse in a loop around the band extrema, with the phase equal in magnitude but opposite in sign at the K and K'-valleys, as required by the time-reversal symmetry[1,12-14]. The Berry phase not only has close connections to the optical selection rules that allow optical generation and detection of the valley-polarized carriers by circularly polarized photons[12-16], but also plays a central role in novel electron dynamics and transport phenomena in TMD and graphene layers, such as the valley Hall effect[4-6,13,17].

In principle the Berry phase, together with other effects from inversion symmetry breaking, can have profound consequences for the wavefunction and energy spectrum of the excited states in two-dimensional (2D) materials. TMD monolayers are known to host strongly bound excitons with a remarkably large exciton binding energy due to enhanced Coulomb interactions in 2D[18-20]. It was recently predicted that the Berry curvature of Bloch states can add an anomalous term to the group velocity of electrons and holes and creates an energy splitting between exciton states with opposite angular momentum[1,3,7-10]. Fig. 1a shows a simplified exciton energy spectrum illustrating the exciton fine structure based on our *ab initio* GW-Bethe-Salpeter equation (GW-BSE) calculations. The $2p_+$ and $2p_-$ exciton states are split in energy with opposite order for the K and K' valleys due to the opposite chirality in the two valleys[7-9,13]. Such novel exciton fine structure, which embodies important wavefunction properties arising from the Bloch band geometry, can strongly modify the intraexcitonic light-matter interactions. Experimental observation of this predicted exciton spectrum, however, has been challenging, because it requires new spectroscopic probe that can distinguish both the momentum valley and the exciton angular momentum.

Here, we report the first observation of the Berry-phase effect in the exciton spectrum of MoSe$_2$ monolayer using intraexciton optical Stark spectroscopy. We demonstrate that the degeneracy between the $2p_\pm$-exciton states is lifted by the Berry phase effect, and enabling a valley-dependent Autler-Townes doublet from strong intraexciton light-matter coupling. We coherently drive the intraexciton transitions using circularly-polarized infrared radiation, which couples the 1s exciton to the $2p_+$ or $2p_-$ states selectively through the pump photon polarization (solid arrowed lines in Fig. 1a). The pump-induced changes in the 1s exciton transition are detected by circularly polarized probes, which selectively measure the K or K'-valley excitons. Independent control of pump and probe photon polarization enables us to distinguish the exciton fine structures in the K and K'-valleys. We determine an energy splitting of 14 meV between the $2p_+$ and $2p_-$ exciton states within a single valley, and this energy splitting changes sign between K and K'-valleys. We determine the 1s-2p transition dipole moment to be 55±6 Debye. This leads to an optical Stark shift that is almost 40 times larger than the interband counterpart[21–23] under the same pump detuning and driving optical field strength. Such strong and valley-dependent intraexciton transitions open-up new pathways for the coherent manipulation of quantum states in 2D semiconducting materials using infrared radiation.

To investigate the fine structure of the excitonic p-manifold, we fabricated a high quality $MoSe_2$ monolayer that is encapsulated in hexagonal boron nitride (hBN) layers using mechanical exfoliation and stacking following Ref. 21. The sandwiched hBN-$MoSe_2$-hBN heterostructure was then transferred to an alumina-coated silver surface (Fig. 1b). The device was kept in vacuum at 77K for all optical measurements. Fig. 1c shows the reflection contrast spectrum of the $MoSe_2$ monolayer, which exhibits a prominent A-exciton absorption feature at energy $E_{1s}$ = 1.627 eV with a full width half maximum (FWHM) of 9 meV. This A-exciton peak arises from the optical transition between the ground state and the lowest energy 1s exciton state in $MoSe_2$ monolayer, which is well-separated from the higher-lying exciton states due to strong Coulomb interactions in TMD monolayers[10,11,20,24].

We use intraexciton optical Stark spectroscopy with helicity-defined pump and probe light to selectively access the $2p_+$ or $2p_-$ exciton states in the K and K'-valleys. As illustrated in Fig. 1a, we drive the 1s-$2p_+$ intraexciton transition coherently with a $\sigma^+$-polarized infrared pump and monitor the photoinduced changes in the 1s exciton absorption at K and K'-valleys with $\sigma^+$ and $\sigma^-$-optical probes, respectively. Quantum-mechanical coupling between the infrared photon field and the 1s-$2p_+$ electronic transition leads to an avoided-crossing behavior that modifies the 1s-exciton state systematically with the changing infrared photon energy, as illustrated in Fig. 1d. Specifically, the $|1s, n\hbar\omega\rangle$ and $|2p_+, (n-1)\hbar\omega\rangle$ states hybridize when driven by the infrared pump in the "dressed atom" picture, where $n$ is the integer number of infrared pump photons at frequency $\omega$. When the infrared photon energy is below (above) the 1s-2p$_+$ resonance, the non-resonant hybridization leads to a decreased (increased) energy for the 1s exciton state. When the infrared photons are resonant with the 1s-2p$_+$ transition, perfect hybridization between $|1s, n\hbar\omega\rangle$ and $|2p_+, (n-1)\hbar\omega\rangle$ states lead to an energy splitting in the 1s exciton absorption. The pump-induced optical Stark shift and splitting of 1s exciton in K and K'-valleys can be detected from the 1s exciton absorption spectrum by using $\sigma^+$ and $\sigma^-$-polarized probe light, respectively. This allows us to explicitly identify the 1s-2p$_+$ intraexciton transition in each valley. Since the infrared pump photon energy is much lower than the transition energy of 1s exciton, our measurement scheme probes only the coherent optical Stark effects without non-coherent contribution from real carrier generation.

Fig. 2a-c show the transient reflection signals for the $\sigma^+$-polarized probe, which measures the photoinduced changes of the K-valley exciton transition upon excitation with $\sigma^+$-infrared pump of different energies ($E_p$). The driving pump has an effective driving intensity ($I_{\text{eff}}$) of 7±1 MW/cm$^2$, which corresponds to a local optical field strength ($E_{\text{eff}}$) of 70±10 kV/cm (see supplementary materials). The colors in Fig. 2a-c represent the pump-induced change of the probe reflectivity $\Delta R$, which is directly proportional to the change of absorption. The positive (negative) $\Delta R$, is proportional to the decrease (increase) of absorption. The horizontal and vertical axes show the probe energy and pump-probe time delay $\tau$, respectively. Strong transient signals are present for pump-probe delay closed to zero and they become negligible at pump-probe delays larger than 500fs. These instantaneous signals confirm the optical responses arise from coherent optical Stark effects. By changing the driving energy of the $\sigma^+$-pump, the optical response varies significantly. Figure 2d-e display the corresponding transient spectra of the K-valley exciton at $\tau$ = 0 ps. At $E_p$ = 120 meV, the absorption of K-valley exciton exhibits a decrease above $E_{1s}$ and an increase below $E_{1s}$, corresponding to a red-shift of the 1s exciton resonance due to the optical Stark effect. The photoinduced responses is opposite for $E_p$ = 170 meV, which is dominated by transition energy blue-shift. The spectrum at $E_p$ = 142 meV, on the other hand, shows an increase of absorption at energies both above and below $E_{1s}$ and a reduction of absorption at $E_{1s}$, which is consistent with an energy splitting of the 1s exciton peak.

The evolution of the 1s exciton absorption in monolayer MoSe$_2$ under coherent infrared driving can be better visualized directly from the optical absorption spectra characterized by the imaginary part of optical susceptibility ($\chi_{im}$) (see supplementary information)[25]. Fig. 3a displays the absorption spectra of the 1s exciton at $\tau$ = 0 ps for both K and K'-valleys driven by $\sigma^+$-polarized infrared radiation of different photon energies. It shows clearly that the 1s exciton transition exhibits avoided-crossing behavior in both valleys, which evolves gradually from energy blueshift to splitting and then to redshift as the pump photon energy is decreased. However, there is an important distinction between the K and K' valley spectra: the resonant coupling between the $\sigma^+$-infrared photons and the 1s-2$p_+$ intraexciton transition, which splits the 1s exciton resonance, occurs at driving photon energy of 142 meV and 128 meV in K and K'-valleys, respectively. It shows that the 1s-2$p_+$ intraexciton transition energy differs by 14 meV for the K and K' valleys. Due to the time-reversal symmetry between K and K'-valleys in MoSe$_2$ monolayer, this observation also indicates that the 2$p_+$ and 2$p_-$ exciton states are non-degenerate and has an energy difference of 14 meV in a single valley. (See Fig. 1a)

We further plot the blue- and red-shifted 1s resonance as a function of the infrared pump photon energy in Fig. 3b. We find that the energy shifts induced by the intraexciton optical Stark effect are almost 40 times larger than its interband counterpart at the same pump intensity and resonance detuning[21–23].

Since the driving photon energy is closed to the 1s-2$p_+$ transition and strongly off-resonant from the interband transition, our observation can be qualitatively understood using a model describing a driven three-state system as illustrated in Fig. 1a and Fig. 1d. Under $\sigma^+$-pump radiation, $|1s, n\hbar\omega\rangle$ hybridizes with the $|2p_+, (n-1)\hbar\omega\rangle$, which can be described by the effective Hamiltonian

$$H_{\text{eff}} = \begin{pmatrix} E_{1s} + i\gamma_{1s} & \frac{V_1}{2} \\ \frac{V_1}{2} & E_{1s} + |E_{1s-2p+}| - E_p + i\gamma_{2p_+} \end{pmatrix}. \tag{1}$$

Here $V_1$ is proportional to the 1s-2$p_+$ intraexciton transition dipole moment $\mu_{1s-2p_+}$ via $V_1 = \mu_{1s-2p_+} E_{\text{eff}}$, where $E_{\text{eff}}$ is the local optical field strength on the sample. $E_p$, $E_{1s}$ and $E_{1s-2p_+}$ denote the pump photon energy, the 1s exciton energy and the 1s-2$p_+$ intra-exciton transition energy, respectively. $\gamma_{1s}$ and $\gamma_{2p_+}$ are the half width at half maximum of the 1s and 2$p_+$ exciton modes, respectively. Direct diagonalization of the effective Hamiltonian yields two new eigenstates $|\alpha\rangle$ and $|\beta\rangle$, which are energetically separated by $V' = \sqrt{V_1^2 + (\Delta + i\gamma_{2p_+} - i\gamma_{1s})^2}$, where $\Delta \equiv E_p - |E_{1s-2p_+}|$ is the detuning energy. The optical absorption of the probe photon to the new eigenstates can then be computed (see supplementary information).

Fig. 3b displays the calculated absorption spectra at different driving energy. The $E_{1s-2p+}$ used in the fitting is 142 meV and 128 meV for K and K' valleys, respectively. From the fitting to the experimental data, we extract the exciton-photon coupling constant $V_1$ of 8 meV, which corresponds to a 1s-2$p_+$ intraexciton transition dipole moment $\mu_{1s-2p_+}$ of 55±6 Debye. The $\mu_{1s-2p_+}$ is almost 6 times larger than that effective dipole moment in the interband exciton optical Stark effect[21-23], and it leads to a nearly 40 times larger optical Stark shift under the same driving intensity and detuning.

To better understand the experimental results, we performed *ab initio* GW-BSE calculations using the BerkeleyGW[26-28] package to determine the exciton energy levels and optical selection rules of exciton and intraexciton transitions in monolayer MoSe$_2$. In these calculations, environmental screening effects from the hexagonal boron nitride (hBN) encapsulation layers are included[18] from first-principles (see supplementary information). The simulation confirms the energy level diagram of the 1s, 2$p_+$, and 2$p_-$ excitons and the optical selection rules in K and K'- valleys in Fig. 1a. Our calculations find that the energies of the 1s and 2$p_-$ exciton states are separated by 117 meV, with 2$p_+$ exciton states further separated by 7 meV in K-valley. The energetic order of 2$p_+$, and 2$p_-$ excitons states is opposite in the K'-valley, as a result of time-reversal symmetry. Although the 2$p_\pm$ excitons are dark in linear optics, they are optically active when coupled to the 1s exciton with circularly-polarized light (Fig. 1a). For example, our calculations show that the 1s-2$p_+$ intraexciton transition couple exclusively to the left-handed circularly polarized (denoted as $\sigma^+$) light with a transition dipole moment of 42 Debye. The 1s-2$p_-$ intraexciton transition, on the other hand, coupled exclusively to the right-handed circularly polarized (denoted as

$\sigma^-$) light. The experimentally observed intraexciton dipole moment and valley-dependent exciton fine structure match reasonably well with the *ab initio* GW-BSE calculations.

The combination of $2p_\pm$-exciton splitting and extremely strong intraexcitonic light-matter interaction allow us to observe valley-dependent Autler-Townes doublets at higher pump intensity in MoSe$_2$ monolayer. Towards this goal, we fabricated a hBN-encapsulated MoSe$_2$ heterostructure on a zinc-sulphide (ZnS) substrate, where the local field factor on the sample for the infrared pump light is more favorable than that for MoSe$_2$ on alumina coated silver substrate (supplementary information). In this device the $1s$-$2p_+$ intraexciton transition energies for the K and K'-valleys are determined to be 150 meV and 138 meV, respectively (see supplementary information). Fig. 4a shows the absorption spectra of the 1s exciton at $\tau$ = 0 ps for K and K'-valleys under series of excitation intensity. The $\sigma^+$-pump driving energy is set to 150 meV, which is on resonant with the $1s$-$2p_+$ transition in the K-valley but positively detuned from the $1s$-$2p_+$ transition in the K'-valley. We observe contrasting coherent phenomena between the K and K'-valleys: 1s exciton transition exhibits a striking splitting into the Autler-Townes doublet in the K-valley, but shows a mostly blueshift in the K'-valley. On the other hand, when the $\sigma^+$-pump energy is tuned to 138 meV, which is negatively detuned from the $1s$-$2p_+$ transition in the K-valley but on resonant with the $1s$-$2p_+$ transition in K'-valley, the 1s exciton transition shows red shift in the K-valley but a clear Autler-Townes doublet in the K'-valley (Fig. 4b). Fig. 4c,d show the splitting energy in the Autler-Townes doublet at resonant excitation scales linearly with the excitation field strength, as expected from Eq. 1[29,30]. At an effective driving intensity of 50±10 MW/cm$^2$, which corresponds to a local optical field strength of 200±20 kV/cm, the Autler-Townes splitting can reach ~24 meV in both valleys. This Autler-Townes doublet leads to a valley-dependent electromagnetically induced transparency in the 1s exciton transition, where the absorption at the 1s exciton resonance is reduced by more than 10-fold compared to the undriven exciton (Fig. 4a, b). Our findings offer a new and effective pathway to coherently manipulate the quantum states and excitonic excitations using infrared radiation coupled to the $1s$-$2p_+$ intraexciton transition.

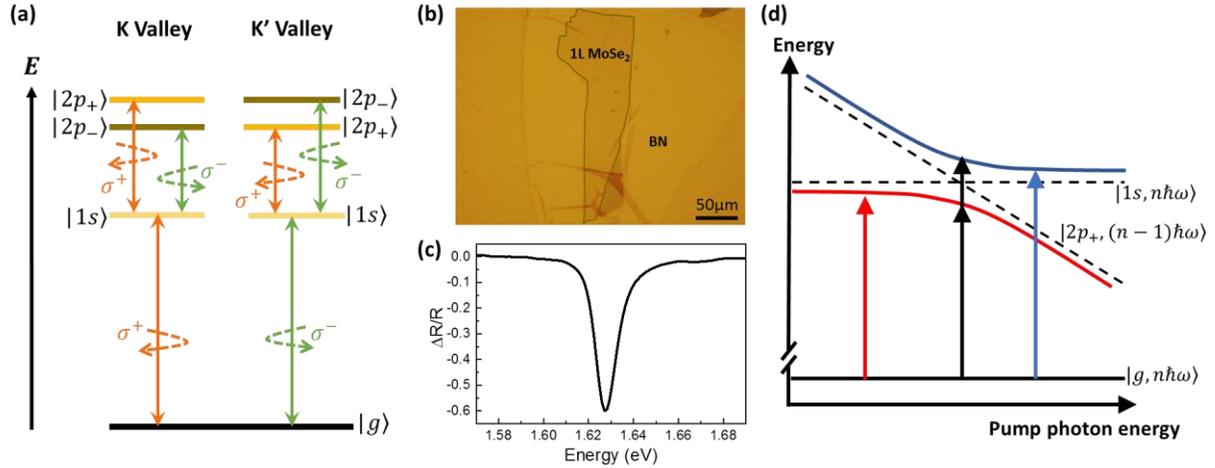

**Fig. 1. Schematics of exciton spectrum and optical transition in MoSe$_2$ monolayer. a.** Illustration of the optical transition and selection rules for one-photon and two-photon excitations in the K and K' valleys of MoSe$_2$ monolayer. $|g\rangle$, $|1s\rangle$, $|2p_-\rangle$, and $|2p_+\rangle$ denote the ground state, $1s$, $2p_-$, and $2p_+$-exciton states, respectively. The symbol $\sigma^+$ and $\sigma^-$ denotes left and right circular polarization state, respectively. **b.** Optical micrograph of monolayer MoSe$_2$ encapsulated by hBN layers on alumina coated silver substrate. The scale bar corresponds to 50 $\mu$m. **c.** The reflection contrast of hBN encapsulated MoSe$_2$ monolayer on alumina coated silver surface at 77 K. It shows prominent A-exciton resonance at 1.627 eV with a FWHM of ≈9 meV. **d.** Schematic diagram illustrating the avoided-crossing behavior due to quantum-mechanical coupling between the infrared photons field and the 1s-2p$_+$ electronic transition. The dashed lines show the unperturbed $|1s, n\hbar\omega\rangle$ and the $|2p_+, (n-1)\hbar\omega\rangle$ states as a function of the pump photon energy, and the blue and red solid lines show the dressed exciton states from quantum hybridization. The arrows show the optical transitions from the ground state to the dressed 1s exciton state.

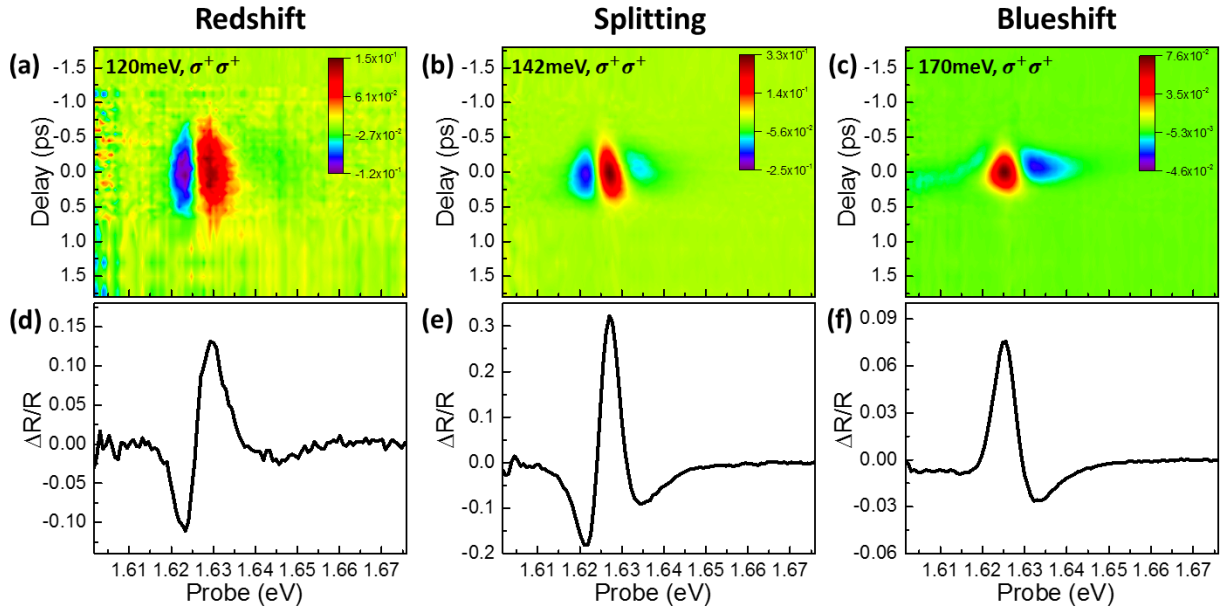

**Fig. 2. Transient reflection spectra of K-valley exciton transitions. a-c.** Two-dimensional plot of transient reflection spectra of the K-valley 1s-exciton resonance of MoSe$_2$ at 77 K following photoexcitation with $\sigma^+$-polarized infrared pump at photon energy of (a) 120 meV, (b) 142 meV and (c) 170 meV. The color scale, vertical axis and horizontal axis represent the relative reflectivity change ΔR/R, the pump-probe time delay $\tau$, and the probe photon energy, respectively. The positive (negative) ΔR/R represents decrease (increase) of absorption. The photoinduced absorption in the K-valley 1s-exciton is monitored by $\sigma^+$-polarized probes. The signals are finite only when the pump and probe pulses overlap in time, indicating an instantaneous coherent response and negligible excitation of real exciton population. **d-f.** At $\tau$ = 0 ps, the coherent signals for $\sigma^+$-probes exhibit spectral responses that are characteristic of (d) energy redshift to (e) splitting and then to (f) energy blueshift as the driving photon energy is increased from 120 meV to 170 meV.

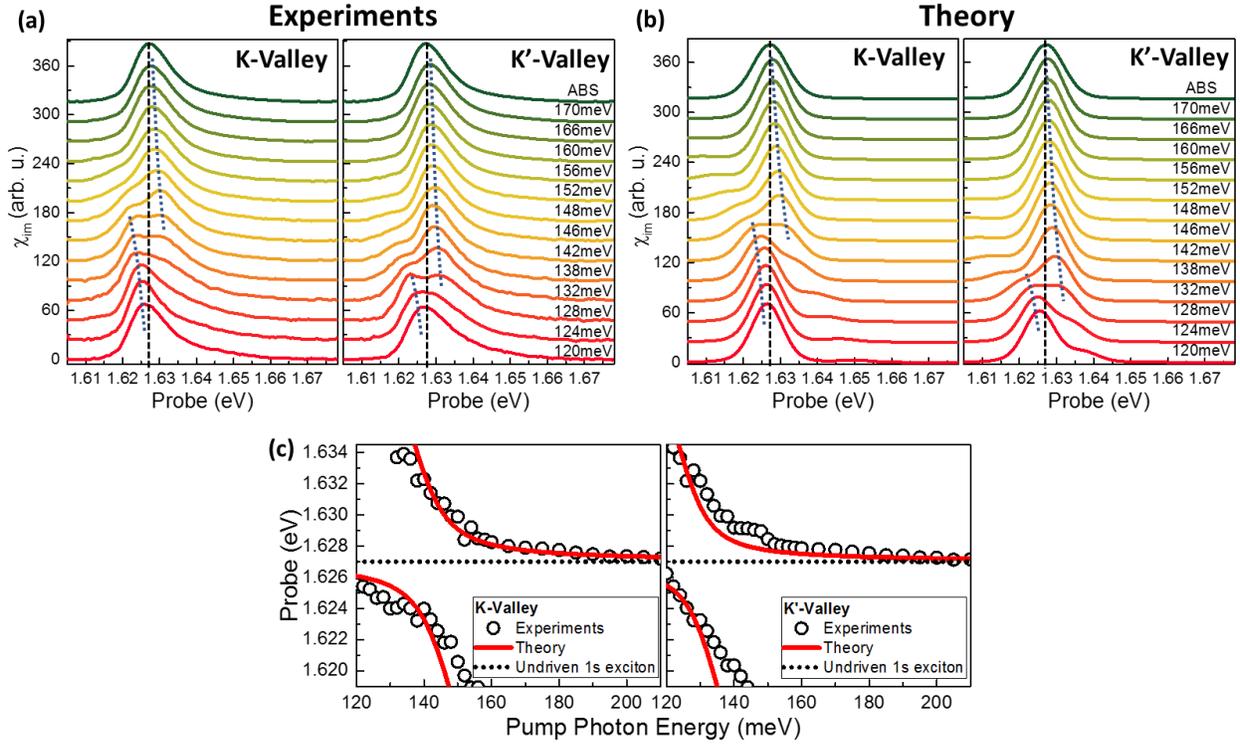

**Fig. 3. Valley-dependent intraexciton optical Stark effect. a-b.** Experimentally observed (a) and calculated (b) photoinduced absorption spectra of MoSe$_2$ monolayer at $\tau$ = 0 ps under various $\sigma^+$- pump excitation energy for the K and K' valleys. The dashed-lines indicate the peak position of unperturbed A-exciton. The dotted lines are guides to the eyes for the peak position at different driving energies. The spectra are offset for clarity and labelled according to the excitation energy (meV). The spectra evolve from energy redshift to splitting and then to blueshift, as the driving energy is increased. The calculation is based on the Hamiltonian shown in Eq. 1. Exciton-photon coupling leads to avoided-crossing and the observed peak splitting at resonant coupling. This resonant coupling occurs at driving photon energy of 142 meV and 128 meV in the K and K' valleys, respectively. It corresponds to a Berry-phase induced 1s-2p$_+$ intraexciton transition energy difference of 14 meV. **c.** Measured 1s exciton peak position (circles) in the K and K' valleys as a function of $\sigma^+$-infrared pump photon energy for an effective driving intensity is 7±1 MW/cm². The solid lines show the calculated dressed-exciton states based on the Hamiltonian shown in Eq. 1.

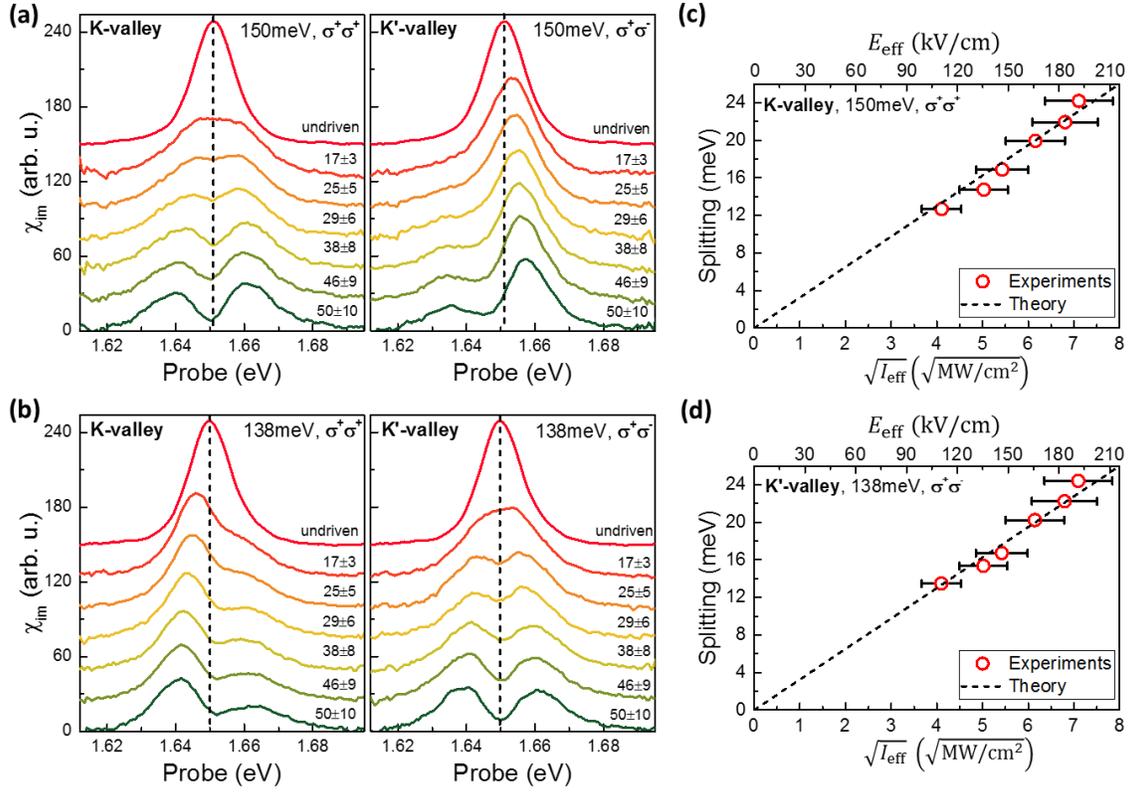

**Fig. 4. Valley-dependent Autler-Townes splitting. a-b.** Photoinduced absorption spectra of MoSe$_2$ monolayer on ZnS substrate at $\tau$ = 0 ps for a series of $\sigma^+$-pump intensity at driving energy of (a) 150meV and (b) 138 meV for K and K'-valley. The dashed-lines indicate peak position of undriven A-exciton. The spectra are offset and labelled according to the effective driving intensity (MW/cm$^2$). **c-d.** The dependence of Autler-Townes splitting energy in K and K'-valley on the squared root of effective pump intensity ($\sqrt{I_{\text{eff}}}$) for driving photon energy of (c) 150 meV and (d) 138 meV, respectively. The top axis shows the corresponding effective local optical field strength ($E_{\text{eff}}$). The splitting energies are obtained from fitting the photoinduced absorption lines with two Lorentzian lines. The dashed lines are fitting obtained from the Hamiltonian shown in Eq. 1.

# Materials and Methods

**Sample Fabrication.** The MoSe$_2$ monolayer encapsulated in hBN flakes were prepared with a polyethylene terephthalate (PET) stamp by a dry transfer method[21]. Monolayer MoSe$_2$ and hBN flakes were first exfoliated onto silicon substrate with a 90 nm oxide layer. We used PET stamp to pick-up the top hBN flake, monolayer MoSe$_2$, and bottom hBN flake in sequence with accurate alignment based on an optical microscope. The hBN/MoSe$_2$/hBN heterostructure was then stamped on a silver substrate coated with a 85 nm alumina layer or on a zinc sulphide (ZnS) substrate. Polymer and samples were heated to 60°C for the pick-up and 130°C for the stamping process. Finally, the PET was dissolved in dichloromethane for 12 hours at room temperature. The sample temperature was kept at 77 K in a liquid-nitrogen cooled cryostat equipped with BaF$_2$ window during optical measurements.

**Intraexciton Optical Stark Spectroscopy.** Pump-probe spectroscopy study is based on a regenerative amplifier seed by a mode-locked oscillator (Light Conversion PHAROS). The regenerative amplifier delivers femotosecond pulses at a repetition rate of 150 kHz and a pulse duration of 250 fs, which were split into two beams. One beam was used to pump an optical parametric amplifier and the other beam was focused onto a sapphire crystal to generate supercontinuum light (500-1100 nm) for probe pulses. Femtosecond mid-infrared pump pulses with tunable photon energies were generated via difference frequency mixing of the idler pulses from the optical parametric amplifier and residual of fundamental output (1026 nm) from regenerative amplifier in a 1 mm thick silver gallium sulphide (AGS) crystal. The mid-infrared pulse duration is ~350 fs. The pump-probe time delay was controlled by a motorized delay stage. The probe light was detected by high sensitivity CCD line camera operated at 1000 Hz. The helicity of pump and probe pulses was independently controlled using Fresnel rhomb and broadband quarter-waveplates, respectively. The experiment followed a reflection configuration with a normal incidence and collinear pump-probe geometry, where the absorption spectra are extracted from the reflectance contrast as described in the supporting information.

**Acknowledgments:** This work was primarily supported by the Center for Computational Study of Excited State Phenomena in Energy Materials, which is funded by the U.S. Department of Energy, Office of Science, Basic Energy Sciences, Materials Sciences and Engineering Division under Contract No. DE-AC02-05CH11231, as part of the Computational Materials Sciences Program which provided the experimental measurements and GW-BSE calculations. The sample fabrication and linear optical spectroscopy was supported by the US Army Research Office under MURI award W911NF-17-1-0312. The pump-probe setup was supported by the ARO MURI award W911NF- 15-1-0447. This research used resources of the National Energy Research Scientific Computing Center (NERSC), a DOE Office of Science User Facility supported by the Office of Science of the U.S. Department of Energy under Contract No. DE-AC02-05CH11231, and the Extreme Science and Engineering Discovery Environment (XSEDE), which is supported by National Science Foundation grant number ACI-1548562. S.T. acknowledges support from NSF DMR-1552220. K.W. and T.T. acknowledge support from the Elemental Strategy Initiative conducted by the MEXT, Japan and the CREST (JPMJCR15F3), JST. E.C.R acknowledges support from the Department of Defense (DoD) through the National Defense Science & Engineering Graduate Fellowship (NDSEG) Program. C.-K.Y. and C.S.O. acknowledges useful discussion with Prof. Ajit Srivastava.

**Author contributions:** C.-K.Y. and F.W. conceived the project. C.-K.Y. designed the experiments and carried out optical measurements and assisted by J.H.. C.-K.Y., F.W. and analyzed the data and performed theoretical analysis assisted by M.I.B.U. M.I.B.U., C.-K.Y., E.C.R. and A.Z. fabricated the devices. C.S.O., T.C. and S.G.L. performed GW-BSE calculations, Y.S, H.C. and S.T. synthesized $MoSe_2$ crystals. K.W. and T.T. synthesized hBN crystals. C.-K.Y. and F.W. wrote the manuscript with inputs from all authors.

**Competing interests:** Authors declare no competing interests.

**Supplementary Information:** Supplementary information to the text can be provided upon reasonable request.